\documentclass[aps,pra,twocolumn,superscriptaddress]{revtex4-2}
\usepackage{amsmath}
\usepackage{amsfonts}
\usepackage{bm}
\usepackage{braket}
\usepackage{graphicx}
\usepackage{siunitx}
\usepackage{dsfont}

\usepackage{xcolor}

\usepackage{orcidlink}

\newcommand{\norm}[1]{\left|#1\right|}
\def\mi{\text{i}}
\def\ap{\hat{a}^{\text{p}}}

\def\as{\hat{a}^{\text{s}}}
\def\asd{\hat{a}^{\text{s}\dagger}}
\def\ai{\hat{a}^{\text{i}}}
\def\aid{\hat{a}^{\text{i}\dagger}}
\def\Ham{\hat{H}}

\def\hc{\text{H.c.}}
\def\vp{v^{\text{p}}}
\def\vs{v^{\text{s}}}
\def\vi{v^{\text{i}}}

\def\nup{\nu^{\text{p}}}
\def\md{\mathrm{d}}
\def\dt{\mathrm{d}t}

\def\alphap{\alpha^{\text{p}}}

\newcommand{\commu}[1]{\left[#1\right]}

\def\his{\hat{H}^{\text{is}}}
\def\uis{\hat{U}^{\text{is}}}

\usepackage[]{hyperref}

\begin{document}
\title{Self-induced manipulation of biphoton entanglement in topologically distinct modes}
\author{Wei-Wei Zhang\,\orcidlink{0000-0002-8164-9527}}
% \email{}
\affiliation{%
School of Computer Science, Northwestern Polytechnical University, Xi’an 710129, China 
}
\author{Chao Chen\,\orcidlink{0000-0002-4476-4366}}
 \email{chenchao2@nbu.edu.cn}
\affiliation{%
School of Physical Science and Technology, Ningbo University, Ningbo 315211, China
}

\author{Jizhou Wu\,\orcidlink{0000-0003-4732-1437}}
\email{wuchichou@gmail.com}
\affiliation{Quantum Science Center of Guangdong-Hong Kong-Macao Greater Bay Area (Guangdong), Shenzhen, 518045, China.}
\affiliation{Department of Physics, Southern University of Science and Technology, Shenzhen, 518055, China}

\begin{abstract}
Biphoton states have been promising applications in quantum information processing, including quantum communications, quantum metrology, and quantum imaging. The generation and manipulation of biphoton entanglement in topologically distinct modes paves the way in this direction. Here we present a comprehensive method for regulating the topological properties of the system by combining the nonlinearity in waveguides, i.e. nonlinearity in the waveguide coupling materials, and the waveguide lattice structure. 
Our method enables the generation of topological biphoton states with the injected pump activation on the topologically trivial modes.  This is realized with the self-induced manipulations on pump-dependent nonlinear couplings on the defects, which is unable to be realized while there are no such nonlinear couplings. 
Specifically, by including the nonlinear gain/loss mechanism in the coupling between the nearest neighbor waveguides and the 
third-order Kerr nonlinearity effect along the waveguides, the injected pump power will be the controllable parameter for the manipulation of the topology in the defect states and the generation of biphoton entanglement states. 
We also present an experimental proposal to realize our scheme and its generalization in the contemporaneous ``active'' topological photonics time-bin platforms.  
Our method can be used in other SSH models with various defect configurations. Our method enables the reusability and versatility of SSH lattice chips and their application for fault-tolerant quantum information processing, promoting the industrialization process of quantum technology.

\end{abstract}
\maketitle

\section{Introduction}
The discovery of topological insulators has revolutionized the field of condensed-matter physics,  which exhibits a unique duality in its electronic properties, insulating in bulk yet conducting along its surface or edges~\cite{PhysRevLett.49.405,wen1995topological,RevModPhys.82.3045}. 
The fundamental ideas underlying these topological edge states are not confined to electronic condensed-matter systems. Topological phases can be achieved using other physical platforms as well, among which are electromagnetic systems like photonic crystals and metamaterials~\cite{lu2016topological,ozawa2019topological,xie2018photonics}. 
By exploiting geometrical and
topological ideas to design and control the behavior of light, topological photonics platforms have been used to 
 engineer analogous effects  for
photons~\cite{khanikaev2013photonic,perchikov2023asymmetric,li2021reduced,2023-10.1063/5.0138763,shlivinski2025universal,2025-PhysRevLett.134.116605}. With the flexibility and diversity to change the dimensionality and symmetries of photonics systems, different topological behaviors can be explored~\cite{ozawa2019topological}.

An astonishing property of the topologically protected edge states is that they are immune to system perturbations and impurities, allowing for the ballistic transport of electrons~\cite {susstrunk2015observation}. The engineering of topologically protected edge states is extensively explored in topological photonics~\cite{Blanco-Redondo2016Phys.Rev.Lett.,barik2020chiral,wang2022topologically,on2024programmable}. In addition to the ingenious design of lattice structure, the effect of waveguide nonlinearity in system topology and topological edge states has been investigated~\cite{blanco2018topological} and has been utilized to generate entangled topologically protected distinct lattice mode states and biphoton states~\cite{wang2019topologically,doyle2022biphoton}. In addition to the nonlinearity in waveguides being an ingredient of the engineering of the system topology, nonlinear materials in laterally 
doped regions could introduce nonlinear interactions between waveguides and therefore offer extra mechanics for topology engineering~\cite{hennig1999wave,Hadad2016Phys.Rev.B,smirnova2020nonlinear,lumer2013self,2023-10.1063/5.0138763,
georgiev1992cavity,herrmann1993starting,snyder1994dynamic,lumer2013self}.

In this work, we propose a comprehensive method for regulating the topological properties of a lattice-structured waveguide system by combining the nonlinearity in waveguides and in the coupling materials between the waveguides. In our scheme, the injected pump power 
manipulates
the coupling of nearest-neighbor waveguides via the self-induced nonlinearity. We present a specific case demonstrating the self-induced manipulation of biphoton entanglement in topologically distinct modes in a subtly designed defective Su-Schrieffer-Heeger~(SSH) model realized in waveguide chips. 
In our scheme, with the evolution of external pump photons, the system Hamiltonian is time-dependent, and therefore the corresponding system spectrum and topology are also time-dependent. 

We study the properties of the corresponding topology zero-energy eigenmode and the pump-induced isolated topologically trivial modes. We discover that the overlap between the topological mode and the trivial localized modes plays a vital role in generating topologically entangled biphoton states. With the increase of the pump power, 
the weight of the topological biphoton state increases for a certain range and then drops,
which demonstrates the validity of our method for the manipulation of topological biphotons even when the pump light does not activate
the topological modes at the initial stage.
We also demonstrate the robustness of our method against off-diagonal disorder in the waveguide coupling. 
With the external manipulation mechanics, our method enables the
reusability of silicon waveguide chips and its application for fault-tolerant quantum information
processing, which has promoted the industrialization process of quantum technology.

\section{The model of nonlinear waveguides chip}

In Sec.~\ref{sec:ssh_defect}, we review two defect configurations of the SSH model. 
In Sec.~\ref{sec:nonlinearity},
we review the mechanism of nonlinearity in waveguide chips.
In Sec.~\ref{sec:defects},
we incorporate the nonlinear coupling for waveguides with defects to provide our design for the waveguides. 
We also provide the eigenspectrum analysis for this case.

\subsection{The SSH model with defects}
\label{sec:ssh_defect}
This section introduces the SSH model, its topological properties, and its variants with defects. 

The SSH model was introduced in 1979~\cite{Su1979Phys.Rev.Lett.}, where the authors solved for the model of a one-dimensional chain with alternating hopping strengths.
For a one-dimensional SSH lattice with $2n$ sites, the Hamiltonian can be written as
\begin{align}
    H = \sum_{j = 0}^{n-1}\left(u a_{2j+1}^{\dagger}a_{2j}+\text{H.c.}\right)
    +\sum_{j = 1}^{n-1}\left(va_{2j}^{\dagger}a_{2j-1}+\text{H.c.}\right)\label{eq:ssh}
\end{align}
with ``H.c.'' being the ``Hermitian conjugate''.
The localized edge states appear at both ends of the one-dimensional chain when $u<v$~\cite{Asboth2016SpringerInternational}.
Such localization behavior has a topological origin:
Solving for the bulk Hamiltonian by setting the periodic boundary to Eq.~\eqref{eq:ssh},
i.e., adding term
$va^{\dagger}_{0}a_{2n-1}$,
we can block-diagonalize the Hamiltonian with each block (two-by-two matrix) $H(k)$ corresponding to a specific momentum $k$.
In the basis of the chiral symmetry
\begin{align}
    \Gamma H(k)\Gamma^{\dagger} = -H(k)
\end{align}
with $\Gamma=\sigma_{z}$,
the Hamiltonian can be written as
\begin{align}
    H(k) = \begin{pmatrix}
        0&h^{*}(k)\\
        h(k) & 0
    \end{pmatrix}
\end{align}
The winding number
\begin{align}
    W=\frac{1}{2\pi\mathrm{i}}\int_{-\pi}^{\pi}\mathrm{d}k\frac{\mathrm{d}}{\mathrm{d}k} \ln h(k)
\end{align}
can be used to characterize
the homotopy of the mapping $k \to H(k)$.
With $W = 1$~($W = 0$), i.e., $u<v$~($u>v$), the bulk is topologically nontrivial~(trivial).

Based on the SSH model, we add an extra site to mimic the defect in real space. As per Ref.~\cite{Blanco-Redondo2016Phys.Rev.Lett.}, 
we are concerned with two kinds of defects:
one is the long-long defect that has weaker hopping strengths from the defect to both its adjacent bulks on the left and right; 
the other is the short-short defect that has stronger hopping strengths to both its bulks on the left and right.
Both these two configurations have the zero-energy modes localized around the defect~\cite{Blanco-Redondo2016Phys.Rev.Lett.}.
A distinguishing feature between these two defect configurations is the existence of the trivial localized state in the short-short defect case~\cite{Blanco-Redondo2016Phys.Rev.Lett.},
which can be engineered by the boundary configurations
and will be destroyed by the chiral-symmetric disorder.

Specifically, we have the following for the existence of the edge states:
In the long-long defect case, the chain can be seen as two topologically nontrivial half-infinite SSH chains glued together at the defect site. 
Thus,
only the topological edge state peaks at the defect, which is the edge of an SSH chain in the topological region.
In the short-short defect case,
the glued SSH chains are two topologically trivial half-infinite SSH chains.
Thus, 
if we place the defect at the middle (site 0) in this SSH chain with short-short defect,
no topological edge states exist at the defect~(0) site.
As there are three sites for the defect region,
we can focus one site left~($-1$) and right~($1$) to see its edge behavior.
Thus, the $-1$ (or $1$) site can be seen as the edge of the left~(or right) half-infinite SSH chain with nontrivial topology.
As a result,
we can observe the localization at $\pm 1$ sites for the topological edge states with zero energy, 
and meanwhile the defect-induced 
topologically trivial localized state peaks at the defect.

\subsection{Nonlinearity in waveguide chips}
\label{sec:nonlinearity}
\begin{figure}
  \centering
  \includegraphics[width=0.9\linewidth]{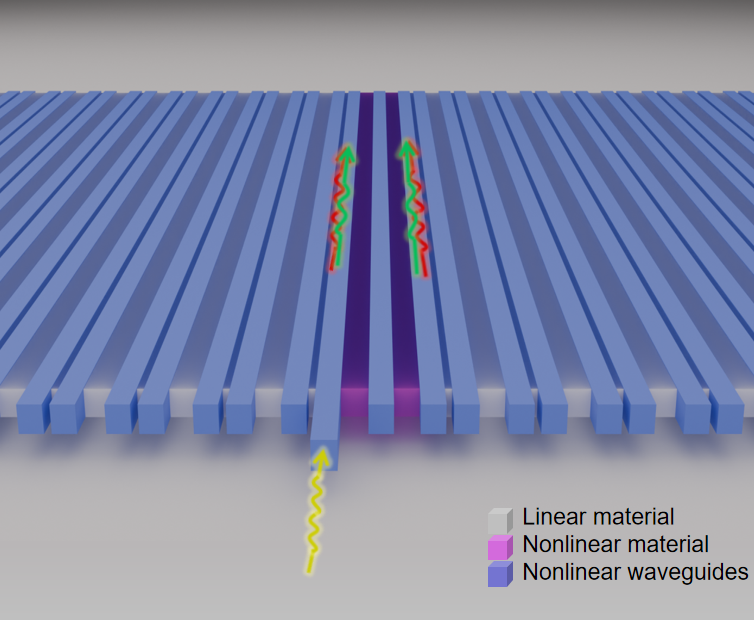}
  \caption{The waveguide design. The pump light is injected at the left side of the center of the waveguides, with the propagation direction indicated by the yellow arrow (indexed as $-1$). The generated biphoton is indicated by the green and red arrows.}
  \label{fig:waveguide_design}
\end{figure}

In this section,
we use the waveguide to simulate the SSH model with defects.
More specifically, we introduce the nonlinearity effect that naturally exists in the waveguide system. We use  Kerr-like nonlinearity for the SSH model~\cite{Hadad2016Phys.Rev.B}, where, in addition to the linear hopping, we also consider a nonlinear contribution determined by the density of the two neighboring sites around the hopping and the spontaneous four-wave mixing procedure~(SFWM) for the generation of biphotons. 
Then for the waveguide array with $2N+1$ sites, and with the defect placed at site 0 as shown in Fig.~\ref{fig:waveguide_design},
the
process can be modeled as a nonlinear process in the Hamiltonian:
\begin{align}
    H_{\text{nonlinear}} = \gamma \sum_{j = -N}^{N}(a_{j}^{\text{p}})^2a_{j}^{\text{s}\dagger}a_{j}^{\text{i}\dagger}+\text{H.c.}
\end{align}
where the superscripts ``i'', ``s'' and ``p'' are short for ``idler'', ``signal'' and ``pump'' light,
and $\gamma$ is the corresponding nonlinear coefficient.
The principle of operation of a passively coupled cavity is
based on a fast nonlinear gain/loss mechanism, which in first order can be described by
$\chi_I=\chi_l-bI$,
with $I$ as 
the photon number density and $b$ as the nonlinear gain/loss
coefficient~\cite{georgiev1992cavity,herrmann1993starting,snyder1994dynamic,lumer2013self}.
The  nonlinearities may induce topological order, and
the temporal dynamics of this process have been shown to take place as a
function of excitation~\cite{lumer2013self,Hadad2016Phys.Rev.B}.

\subsection{Nonlinear coupling in the SSH model with long-long defects} 
\label{sec:defects}

Now, we give our design to manipulate the biphoton generation in waveguides. 
As shown in Fig.~\ref{fig:waveguide_design}, our setup is the SSH model with a long-long defect and extra nonlinear materials~(shown in purple) filled in the neighboring space of the defect.
The filled nonlinear material essentially enables the hopping strength dependence on pump power. 
Therefore, it is desired that, with the adjustment of the injected pump power, the topological structure of the defect could be manipulated,
which offers a flexible and efficient method for the biphoton generation in the waveguide.

The evolution of the pump light is mainly determined by the
nonlinear gain/loss mechanics between waveguides with the gap filled with a nonlinear material, which follows~\cite{PhysRevE.67.056606,10.1063/1.522558,10.1063/1.523009}.
\begin{align}
        \mathrm{i}\frac{\md}{\dt}\alphap_{j}(t) 
        = \vp_{j-1}\alphap_{j-1}(t) +\vp_{j}\alphap_{j+1}(t) 
        \label{eq:Hp}
\end{align}
where
$j\in [-N,N]:=\set{-N,-N+1,\cdots,N}$, and the periodic boundary is imposed at $j=-N$ and $j=N$. Here
$\alphap$ is the amplitude for the pump;
$\vp$ is the hopping strength
with $\vp_{j}=\vp_{\text{long}}$ for $j\in \mathsf{L}:=\set{-3,-5,\cdots}\cup\set{2,4,\cdots}$, which describes the linear hopping between waveguides with a long separation;
$\vp_{j}=\vp_{\text{short}}$ for $j\in \mathsf{S}:=\set{-2,-4,\cdots}\cup\set{1,3,\cdots}$, which gives the linear hopping between waveguides with a short separation; and for $j\in \mathsf{N}:=\set{-1,0}$
\begin{align}
    \vp_{j}=\vp_{\text{long}}+\nup\left(\norm{\alphap_{j+1}(t)}^2+\norm{\alphap_{j}(t)}^2\right)
\end{align}
with $\nup$ the nonlinear coefficient for the hopping between waveguides filled with the nonlinear material.

Based on the pump light evolution, we 
investigate
the evolution of the signal and idler lights. 
Compared with the strong pump light, where we apply 
the
classical approximation,
the idler and signal photons are weak enough so that they need to be treated quantumly. 
The corresponding Hamiltonian is
\begin{align}
    \Ham_{\text{is}} =  \sum_{j=-N}^{N} \left[\vs_{j}\as_{j}\asd_{j+1}
    +\vi_{j}\ai_{j}\aid_{j+1}
    +\gamma\left(\ap_{j}\right)^2\asd_{j}\aid_{j}+\text{H.c.}\right].
    \label{eq:His}
\end{align}
Similar to the setup for the pump light, here 
\begin{align}
    v^{\text{s}/\text{i}}_{j}=\begin{cases}
        v^{\text{s}/\text{i}}_{\text{long}} & j\in \mathsf{L}\\
        v^{\text{s}/\text{i}}_{\text{short}} & j\in \mathsf{S}\\
        v^{\text{s}/\text{i}}_{\text{long}}+\nu^{\text{s}/\text{i}}\left(\norm{\alphap_{j+1}(t)}^2+\norm{\alphap_{j}(t)}^2\right) & j\in \mathsf{N}
    \end{cases}
\end{align}
Using the classical approximation for the pump light,
\begin{align}
    \ap\to\alphap
\end{align}
we have
\begin{align}
    \Ham_{\text{is}} =   \sum_{j=-N}^{N} \left[\vs_{j}\as_{j}\asd_{j+1}
    +\vi_{j}\ai_{j}\aid_{j+1}
    +\gamma\left(\alphap_{j}\right)^2\asd_{j}\aid_{j}+\text{H.c.}\right]
 \end{align}
We decompose this into the hopping part and biphoton generation part from the SFWM process as follows: 
\begin{align}
    \his_{\text{hopping}}&= \sum_{j=-N}^{N} \vs_{j}\as_{j}\asd_{j+1}
    +\vi_{j}\ai_{j}\aid_{j+1} + \hc\\
    \his_{\text{biphoton}}&= \sum_{j=-N}^{N} \gamma\left(\alphap_{j}\right)^2\asd_{j}\aid_{j}+\text{H.c.}
\end{align}

\subsubsection{Hopping part}
The hopping evolution for the generated signal~(or idler) state follows, 
\begin{align}
    \mi \frac{\md \hat{a}^{\text{s}/\text{i}}_{j}}{\dt} = \commu{\hat{a}^{\text{s}/\text{i}}_{j},\his_{\text{hopping}}} = v^{\text{s}/\text{i}}_{j}\hat{a}^{\text{s}/\text{i}}_{j+1} + v^{\text{s}/\text{i}}_{j-1}\hat{a}^{\text{s}/\text{i}}_{j-1} 
\end{align}
with $j\in[-N,N]$.
Here we denote the corresponding matrix~(or the single-particle representation) of the operator $\his_{\text{hopping}}$ on the idler and signal as $H^{\text{i}}$ and $H^{\text{s}}$, respectively.

Consider the dynamics in a short time duration $(t,t+\delta t)$.
The equivalent evolution matrices are denoted as $U^{\text{i}}$ for the idler light and $U^{\text{s}}$ for the signal light, where each $U$ is defined as~$U:=\exp\left(-\mathrm{i}H\delta t\right)$, which transforms the signal and idler lights as
\begin{align}
    \ket{1}_{j}^{\text{i}} \to \sum_{k=-N}^{N}U_{k,j}^{\text{i}}\ket{1}_{k}^{\text{i}}\\
    \ket{1}_{j}^{\text{s}} \to \sum_{k=-N}^{N}U_{k,j}^{\text{s}}\ket{1}_{k}^{\text{s}}
\end{align}
with $j\in[-N,N]$.
For a general two-photon state at time $t$, one has
\begin{align}
    \ket{\Psi(t)} = \sum_{j,k=-N}^{N}m_{j,k}\ket{1}_{j}^{\text{i}}\ket{1}_{k}^{\text{s}}\label{eq:psi-biphoton}.
\end{align}
We can use a matrix $M(t)$ to represent such a state evolved with time $t$ with $M(t)=(m_{i,j})$. Then the
corresponding linear dynamics
\begin{align}
    \ket{\Psi(t)} \to \ket{\Psi(t+\delta t)}=&\sum_{p,q}\sum_{j,k}m_{j,k}U_{p,j}^{\text{i}}U_{q,k}^{\text{s}}\ket{1}_{p}^{\text{i}}\ket{1}_{q}^{\text{s}}\\
    =&\sum_{p,q}\sum_{j,k}m_{j,k}U_{p,j}^{\text{i}}U_{k,q}^{\text{s}\top}\ket{1}_{p}^{\text{i}}\ket{1}_{q}^{\text{s}}
    \label{eq:ketpsitosumpq}
\end{align}
can be described as 
\begin{align}
    M(t)\to M(t+\delta t)=U^{\text{i}}M(t)\left(U^{\text{s}}\right)^{\top}
\end{align}
where ``$\top$'' is used to denote the matrix transpose.

\subsubsection{Biphoton generation part}
The corresponding unitary operator for the nonlinear part in fact gives rise to a squeezelike operator. 
Considering the photon pair generation within the duration $(t,t+\delta t)$ when
$\delta t\ll 1$, we have the following approximation for the nonlinear process:
\begin{align}
    &\uis_{\text{nonlinear}}(t,t+\delta t)\notag\\
    \approx&  \exp\left(-\mi\his_{\text{nonlinear}}(t)\delta t\right)\\
    \approx& \mathds{1} -\mi \left[\sum_{j=1}^{N} \gamma\delta t\left(\alphap_{j}(t)\right)^2\asd_{j}\aid_{j}+\text{H.c.}\right]
\end{align}
which acts on the vacuum state for photon pair generation~(before the pump light propagates to the position):
\begin{align}
    &\uis_{\text{nonlinear}}(t,t+\delta t)\ket{0}\notag\\
    \approx&\left\{\mathds{1} -\mi \left[\sum_{j=1}^{N} \gamma\delta t\left(\alphap_{j}\right)^2\asd_{j}\aid_{j}+\text{H.c.}\right]\right\}\ket{0}\\
    =&\ket{0} -\sum_{j=1}^{N}\mi \gamma\delta t \left(\alphap_{j}\right)^2\ket{1}^{\text{i}}_{j}\ket{1}^{\text{s}}_{j}\label{eq:ket0igammadeltat}
\end{align}
Focusing on the biphoton sector~($\set{\ket{1}^{\mathrm{i}}\!\ket{1}^{\text{s}}}$),
and using the matrix representation of the two-photon state in Eq.~\eqref{eq:psi-biphoton},
the nonlinear process in Eq.~\eqref{eq:ket0igammadeltat}
is equivalent to adding the diagonal terms
to $M$. Then the total transformation on the biphoton state during $(t,t+\delta t)$ with the hopping and biphoton generation process taken into consideration is
\begin{align}
    M(t) \to M(t+\delta t) =& U^{\text{i}}M(t)\left(U^{\text{s}}\right)^{\top} \nonumber\\&-\mi\gamma\delta t\operatorname{diag}\begin{pmatrix}
        \left[\alphap_{1}(t)\right]^2\\
        \left[\alphap_{2}(t)\right]^2\\
        \vdots\\
        \left[\alphap_{N}(t)\right]^2
    \end{pmatrix},
    \label{eq:H-kerr}
\end{align}
which represents the SFWM process along the direction of the waveguides~(the equivalent time direction) for the generation of biphotons.

\begin{figure*}
  \centering
  (a)~\includegraphics[width=0.45\linewidth]{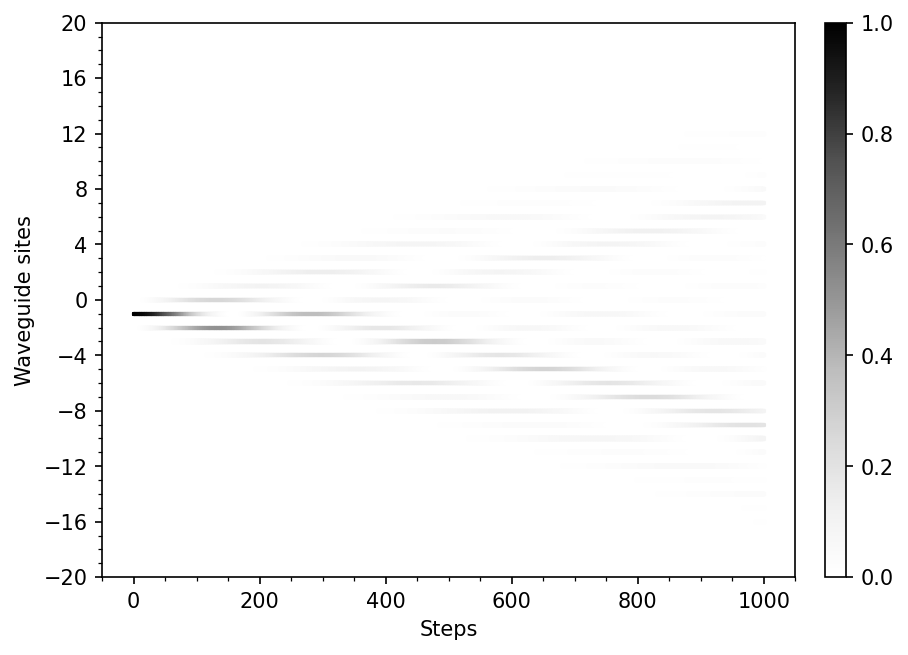}
  (b)~\includegraphics[width=0.45\linewidth]{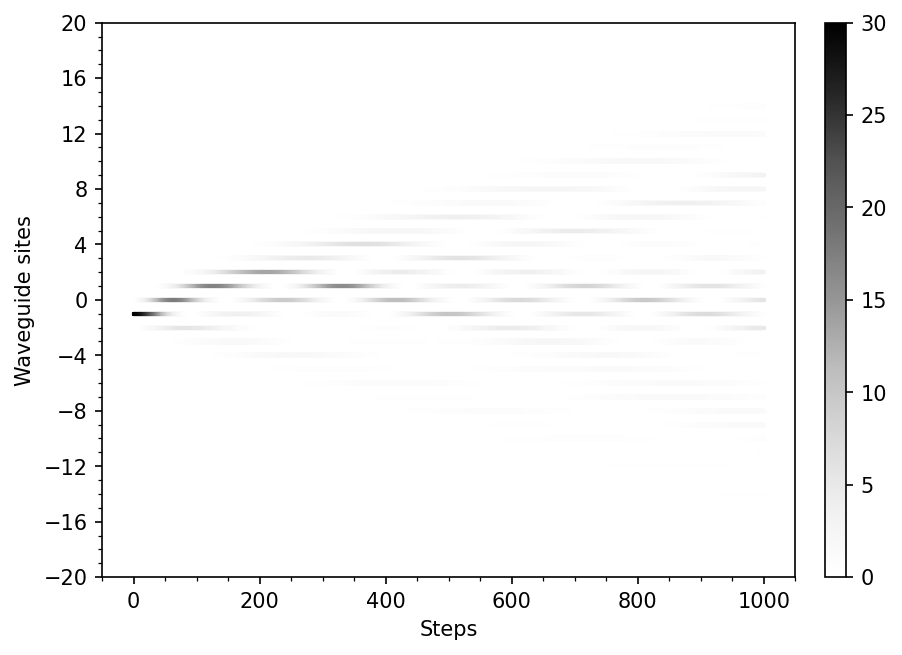}
  (c)~\includegraphics[width=0.45\linewidth]{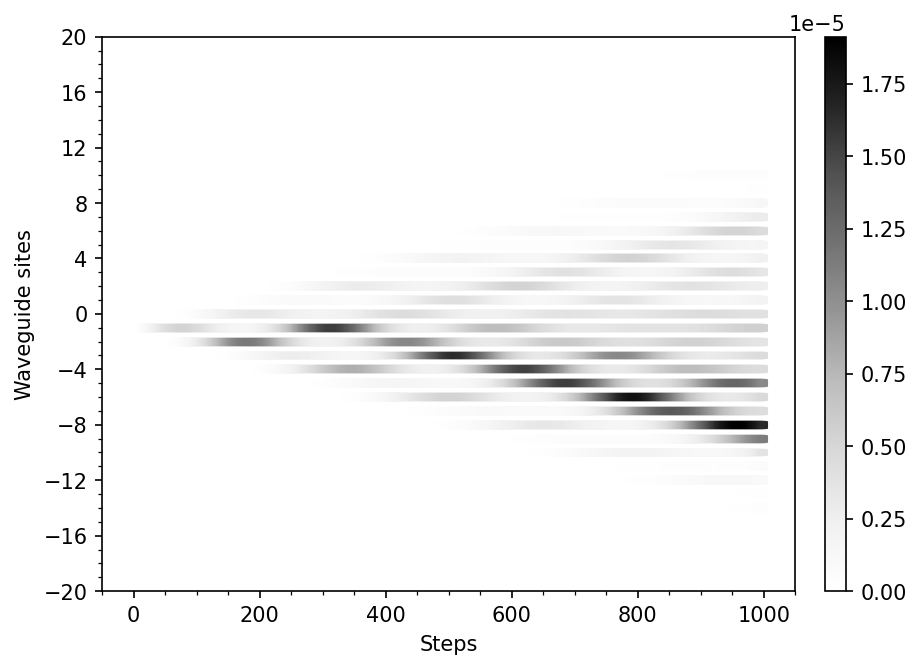}
  (d)~\includegraphics[width=0.45\linewidth]{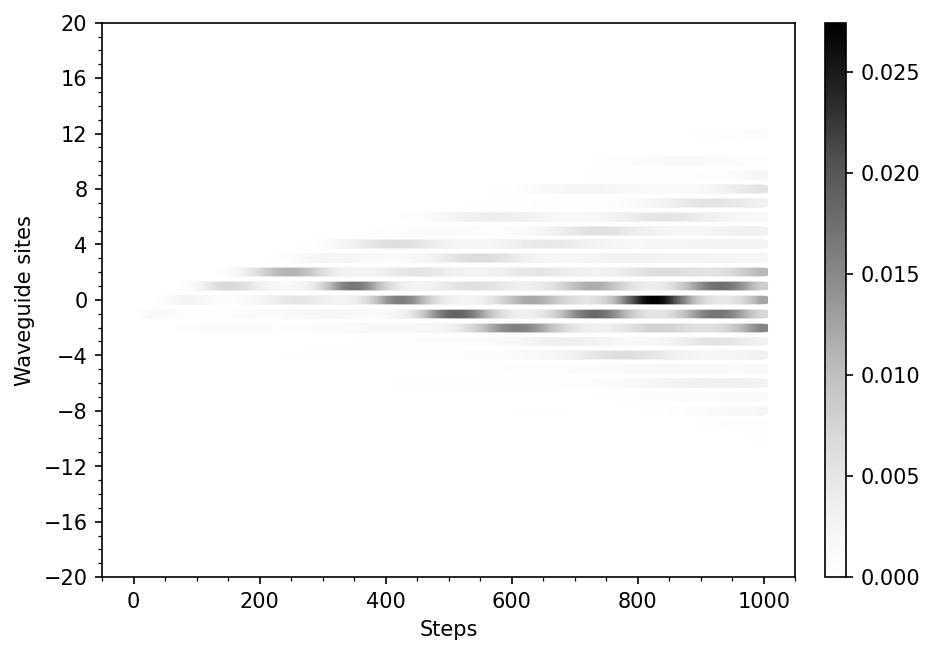}
  \caption{(a),(b)~Pump and (c),(d)~biphoton dynamics in a 103-waveguides chip with a long-long defect centered at the waveguide indexed 0, with different pump power injections: (a),(c)~\qty{1}{\watt} and (b),(d)~\qty{30}{\watt}.
  }
  \label{fig:pump_biphoton_evolution}
\end{figure*}

\section{self-induced manipulation of topological biphoton entangled states}
In this section, we demonstrate the manipulation of the effective topology of the silicon waveguide chip with the external pump power. The structure of our chip architecture is shown in Fig.~\ref{fig:waveguide_design}. There are 103 waveguides in the simulations with the nonlinear gain coefficient $\nup=\nu^\text{s}=\nu^\text{i}=\qty{1078}{\per\metre}$ and linear hopping coefficients $\vp_{\text{long}}=\qty{14951}{\per\metre}$ and $\vp_{\text{short}}=\qty{22118}{\per\metre}$ for the pump photons, $\vs_{\text{long}}=\qty{13603}{\per\metre}$ and $\vs_{\text{short}}=\qty{21882}{\per\metre}$ for the signal photons, and $\vs_{\text{long}}=\qty{14562}{\per\metre}$ and $\vs_{\text{short}}=\qty{22162}{\per\metre}$ for the idler photons.
The nonlinear coefficient for the SFWM is $\gamma=\qty{120}{\per\watt\per\metre}$~\cite{doyle2022biphoton}.
Here the coefficient choice is adopted from previous waveguide experiments~\cite{doyle2022biphoton} with adjustments convenient for the simulations.

\subsection{Self-induced topological pump state}

In our setup, the pump light is injected at the left site (waveguide indexed  $-1$) of the long-long defect as indicated by the yellow arrow in Fig.~\ref{fig:waveguide_design}. 
We investigate the evolution of pump light as pump power increases and discover the localized oscillation of the pump around the defects when the injected pump power is large, as shown in Fig.~\ref{fig:pump_biphoton_evolution}. 
When the pump power is small, the pump light evolves and spreads across the waveguide lattice as time goes on, as shown in Fig.~\ref{fig:pump_biphoton_evolution}(a). 
In the case with a small injected pump power, the evolution dynamics is similar to the case where there are no nonlinear materials around the defect site for the long-long defect case~\cite{Blanco-Redondo2016Phys.Rev.Lett.}.
In this case, the only localized state is the topologically nontrivial edge state with zero energy, which is dominated at the defect (0, $\pm2$) of the waveguides. 
Thus, the injected state has no overlap with the localized state, leading to the spreading behavior.

When we increase the power of the pump light to \qty{30}{\watt}, as shown in Fig.~\ref{fig:pump_biphoton_evolution}(b),
the injected pump has obvious oscillations at the defect and at the $\pm 1, \pm2$ sites.
This is an indication that the system is transformed from the long-long defect case to the pump-induced novel defect case, which means the injected pump power has caused the waveguide chip's defect topology to change. 
In such a case, there are three eigenstates: one is a zero-energy topological edge state, and the other two are topologically trivial edge states,
as shown later in
Figs.~\ref{fig:eigenstate_longlong_pump}(e) and~\ref{fig:eigenstate_longlong_pump}(f).
Different from the short-short defect case analyzed
in Sec.~\ref{sec:ssh_defect},
where the distributions of both
the topological zero-energy state and the defect-induced trivial localized states 
are significant at sites $-1$ and $1$ around the defect,
the dominant of topological state is shifted to the sites $-2$ and $0$ in our pump-induced defect case. 
With the trivial localized states as a bridge,
the pump state has nontrivial overlap with the topological state during the evolution.
Therefore, with the SFWM during the pump evolutions, we can observe a non-negligible oscillation behavior among the defect sites for the population of the generated biphoton states, as shown in Fig.~\ref{fig:pump_biphoton_evolution}(d).

\begin{figure}
  \centering
\includegraphics[width=0.96\linewidth]{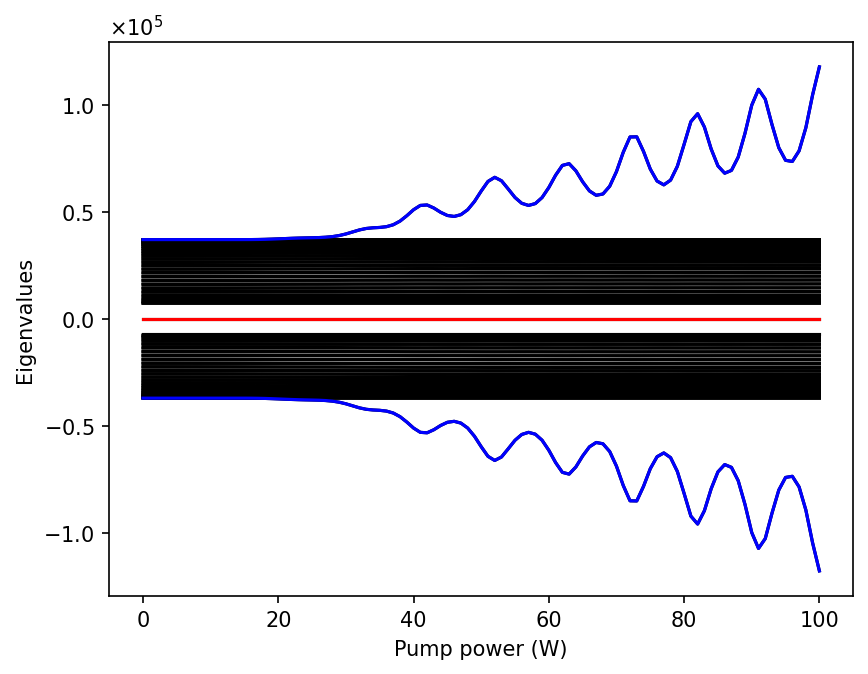}
  \caption{The pump spectrum at the last step of the evolution, where the pump is injected into the waveguide indexed $-1$ of a 103-waveguides chip with a long-long defect centered at the waveguide indexed 0.}   \label{fig:pump_spectrum_versus_pump_power}
\end{figure}

This can be further confirmed from the energy spectrum of the pump light, as shown in Fig.~\ref{fig:pump_spectrum_versus_pump_power}. With the evolution of the pump in our setup, the nonlinearity of the SFWM along the waveguides' direction sets the system Hamiltonian as time-dependent, as the pump population on the waveguides' lattice varies over time. 
When the pump light is set at $\qty{1}{\watt}$, there is only one isolated zero energy, and its eigenstate  is localized at the defect and decays exponentially into the bulk
and the highest~(the lowest) energy state is the bulk state, and no localization behavior is manifested, as shown in Figs.~\ref{fig:eigenstate_longlong_pump}(a)--\ref{fig:eigenstate_longlong_pump}(c). 
When the pump light is increased,
two extra isolated eigenenergies are shown in the spectrum, as shown in Fig.~\ref{fig:pump_spectrum_versus_pump_power}.
These topologically trivial energy-isolated eigenstates are dominated at the central three~(or more) sites depending on the pump distribution, as these are now defects caused by the increase of the effective hopping strength due to the pump power, as shown in Figs.~\ref{fig:eigenstate_longlong_pump}(d)--\ref{fig:eigenstate_longlong_pump}(i).

\subsection{Self-induced topological biphoton states}

With the same setup as in the previous section, we consider both the nonlinear gain and SFWM nonlinear mechanics for investigating the generation of biphotons after the injection of the pump. Without the nonlinear gain, our setup is pump-power-independent and the injection of pump at the $-1$ location lattice of the long-long defect could not generate the topological mode of either pump or biphoton photons. 

When including nonlinear gain, the topology of the waveguide is pump-power-dependent, and, as the pump power increases, 
the long-long defect will be converted into novel types of defects. 
The injection power at $-1$ lattice waveguide will activate the emergent trivial localized mode of the novel defect. 
We discover that, with the SFWM mechanics, the evolution of the trivial localized pump model will generate biphotons in topological modes. 
As shown in Fig.~\ref{fig:pump_biphoton_evolution}(d),
the biphotons are generated following the path of the pump light.
When the pump light is weak 
[\qty{1}{\watt} as shown in Fig.~\ref{fig:pump_biphoton_evolution}(c)],
the biphotons evolve diffusively along the waveguides.
When the pump light is strong~[\qty{30}{\watt} as shown in Fig.~\ref{fig:pump_biphoton_evolution}(d)], the biphotons 
oscillate around the defect.

\begin{figure*}
  \centering
(a)~\includegraphics[width=0.45\linewidth]{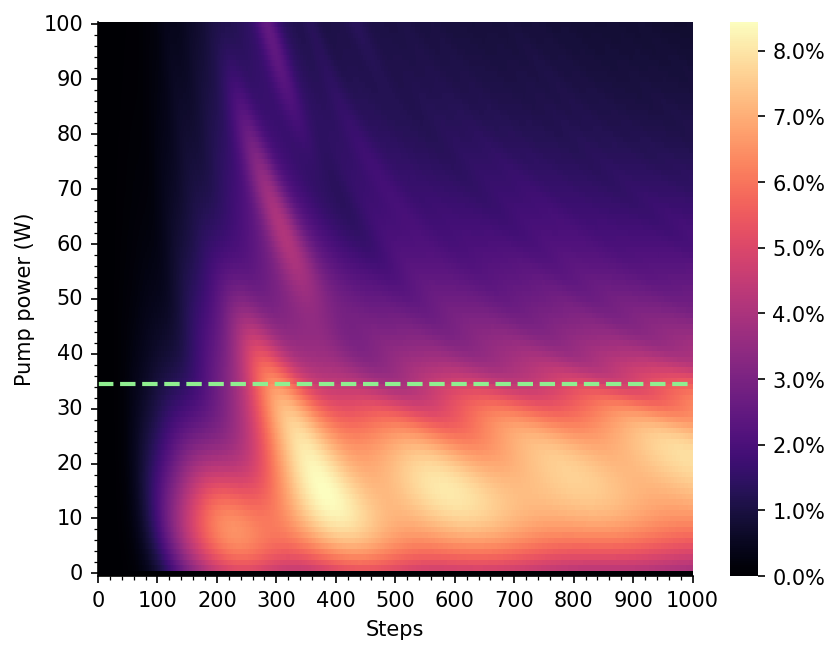}
(b)~\includegraphics[width=0.43\linewidth]{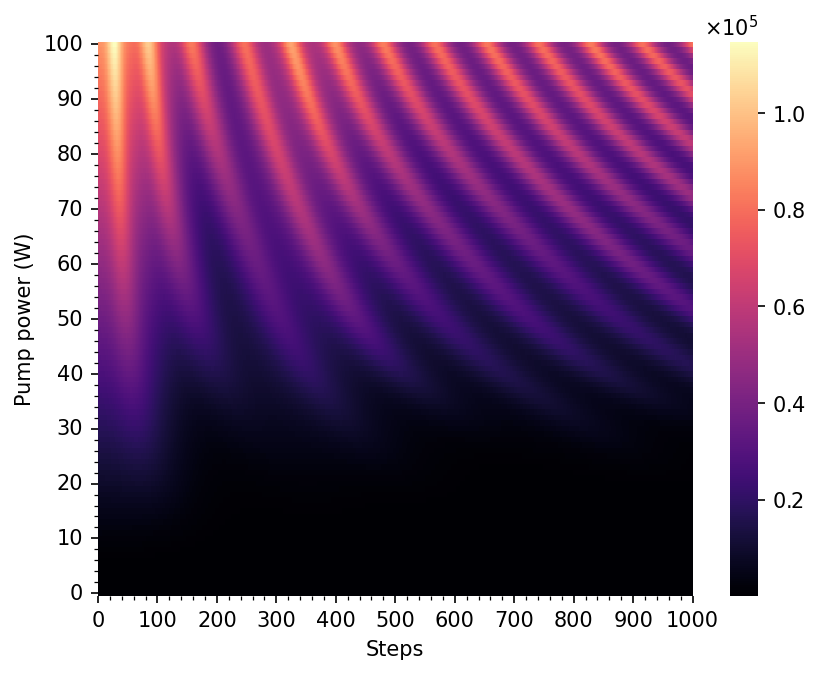}
  \caption{(a)~Heat map of the topological biphoton weight in the generated biphoton states concerning the pump power and evolution steps. (b)~Heat map of the gap between the maximum eigenvalue and the bulk band in the time-dependent pump Hamiltonian spectrum with respect to the pump power and evolution steps. 
  }
   \label{fig:heatmap-weight-gap}
\end{figure*}

With the increase of the pump power, the
process of the defect topology transition can be witnessed by the weight of topological entangled biphoton states in the generated biphoton states, as shown in Fig.~\ref{fig:heatmap-weight-gap}(a). We discover that, as the pump increases from 0 to \qty{100}{\watt}, the weight of topological entangled biphoton states increases first with the increase of pump power, then drops. The increase behavior is consistent with the heat map of the spectrum gap between the topological trivial maximum eigenmode and the bulk band, as shown in Fig.~\ref{fig:heatmap-weight-gap}(b). We discuss the drop behavior later in this section.

\begin{figure*}
  \centering
  (a)~\includegraphics[width=0.3\linewidth]{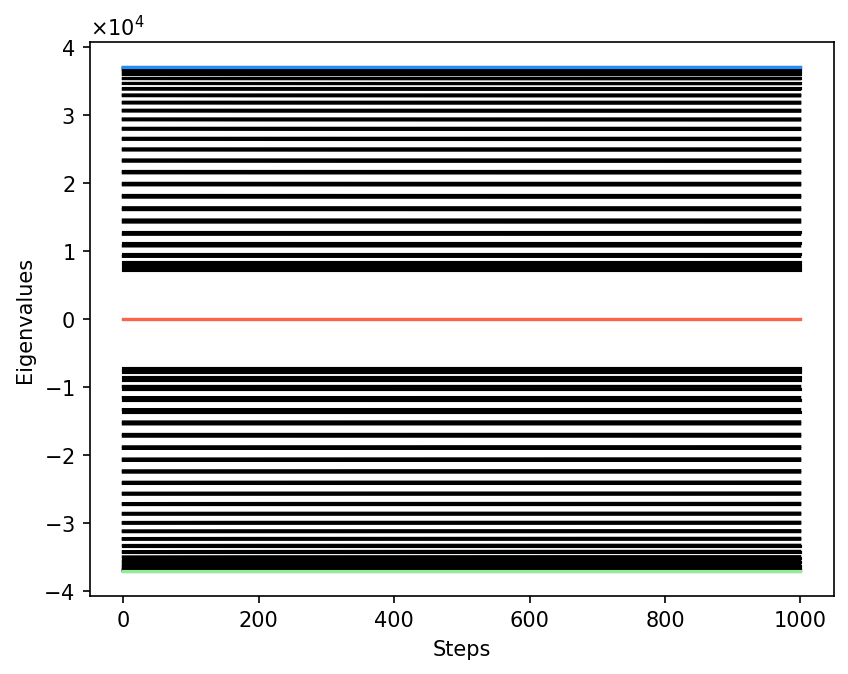}
  (b)~\includegraphics[width=0.3\linewidth]{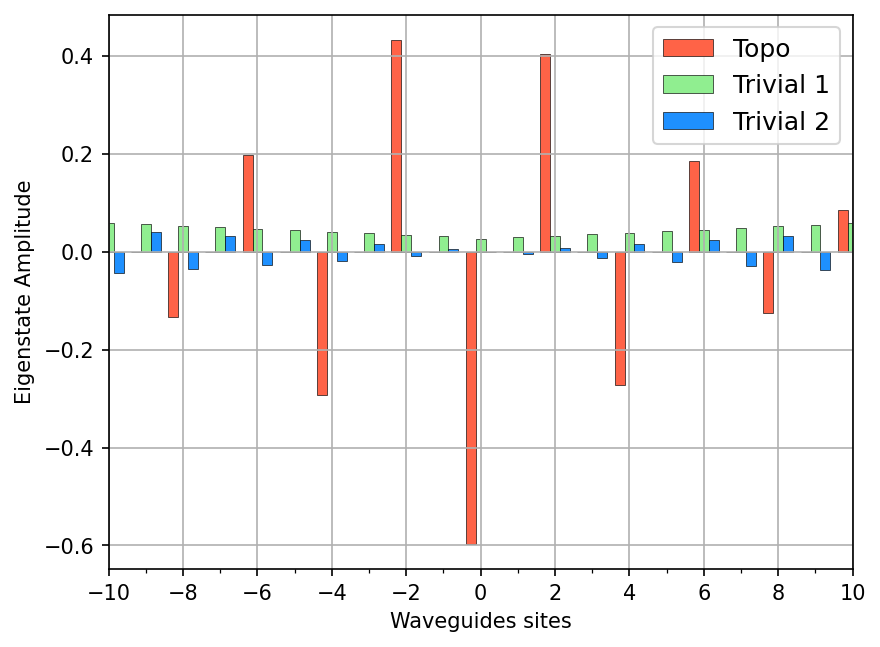}
  (c)~\includegraphics[width=0.3\linewidth]{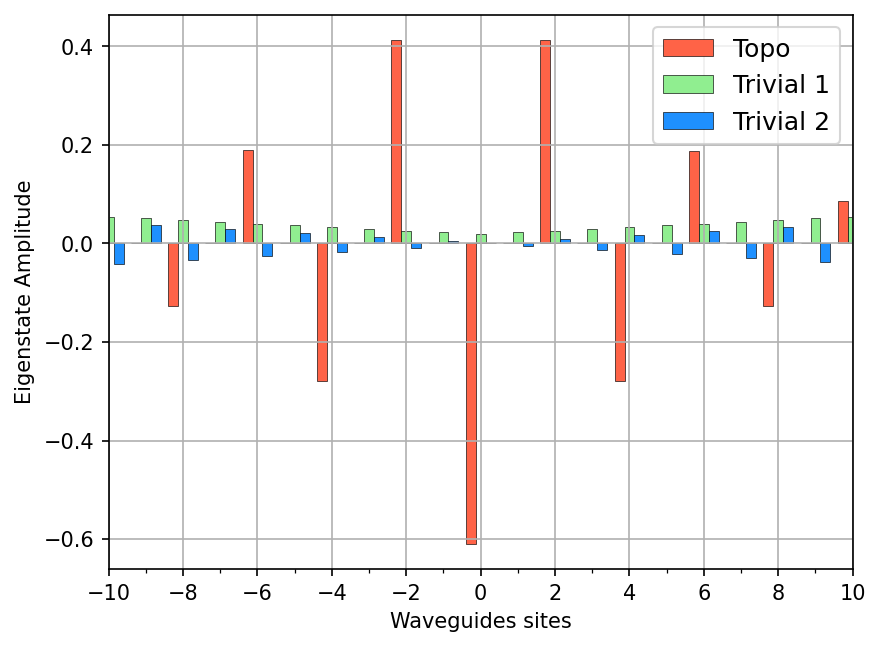}\\
(d)~\includegraphics[width=0.3\linewidth]{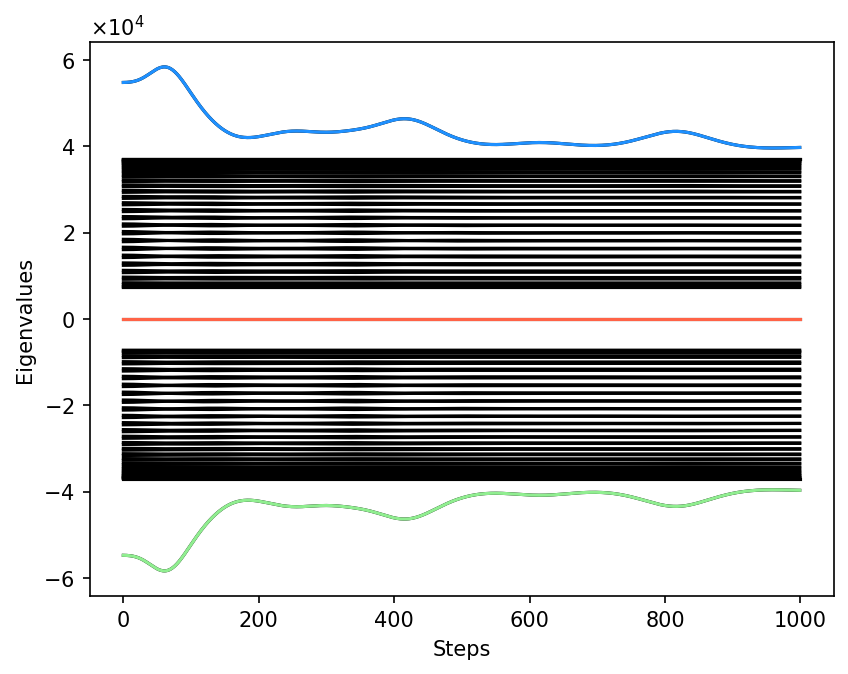}
  (e)~\includegraphics[width=0.3\linewidth]{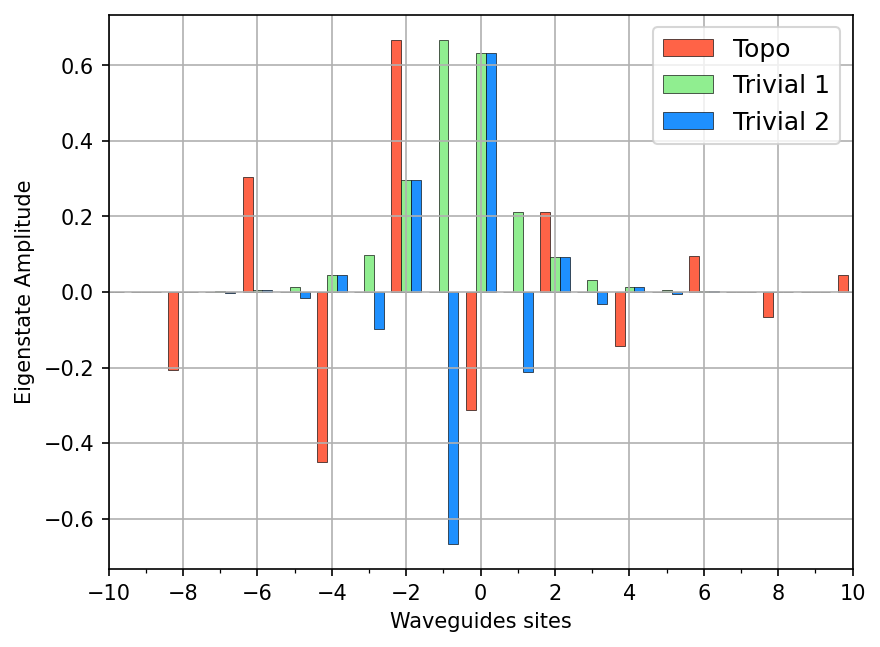}
  (f)~\includegraphics[width=0.3\linewidth]{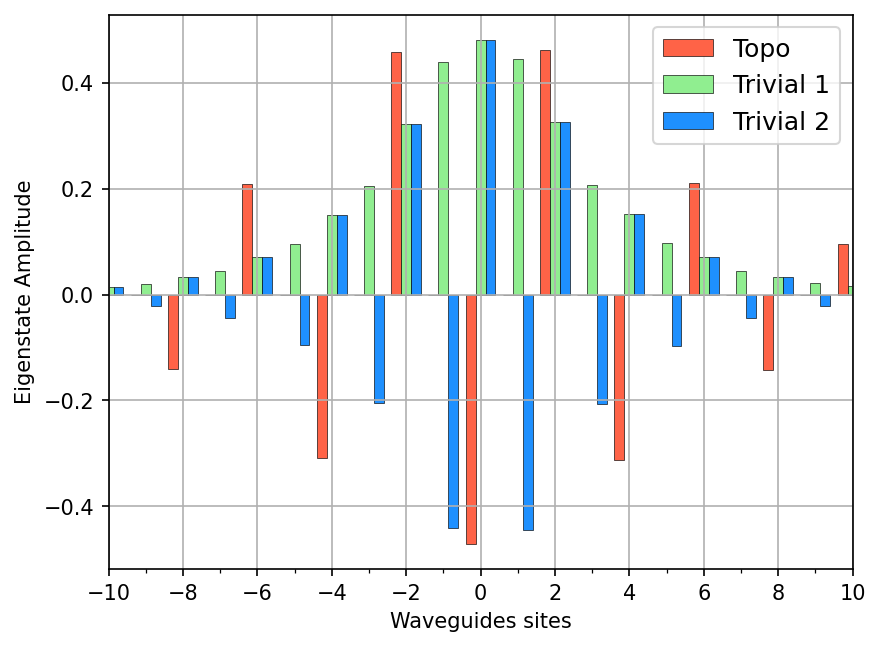} \\   
  (g)~\includegraphics[width=0.3\linewidth]{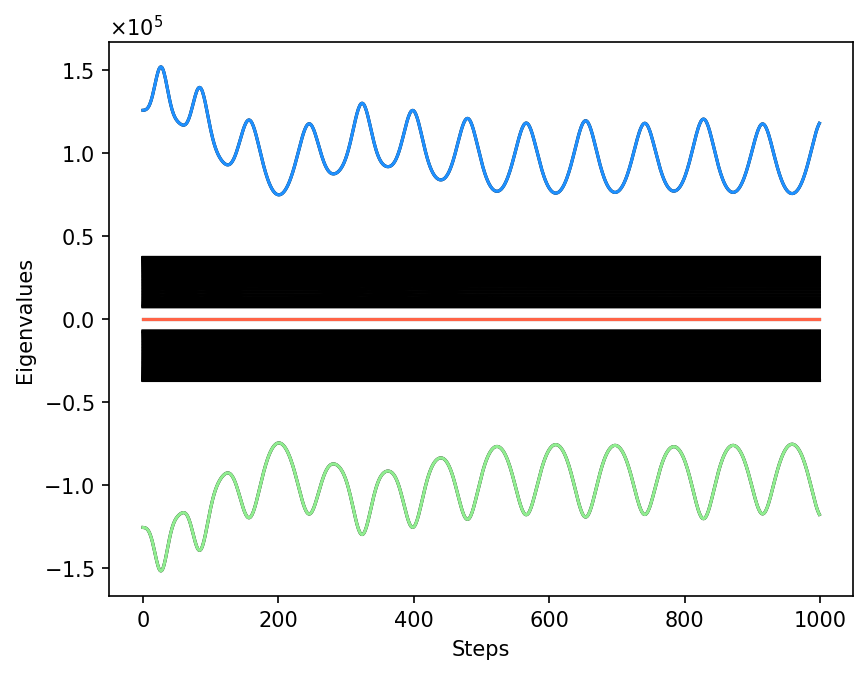}
  (h)~\includegraphics[width=0.3\linewidth]{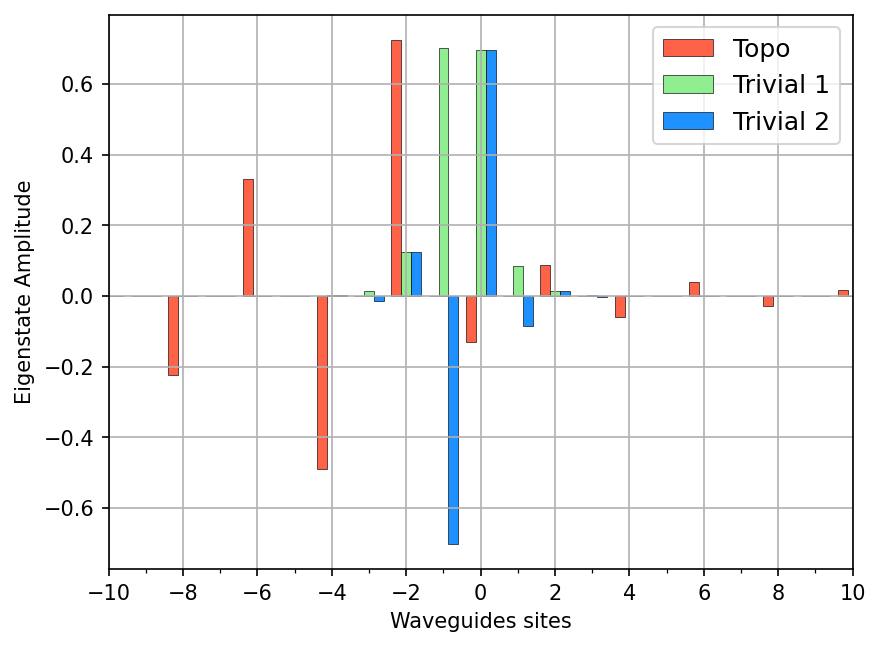}
  (i)~\includegraphics[width=0.3\linewidth]{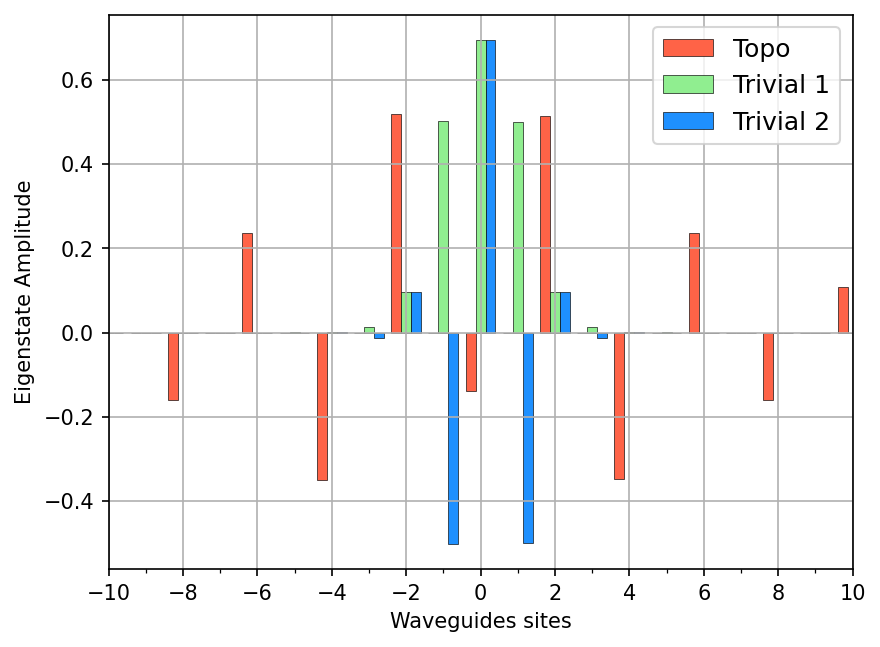}
  \caption{Eigenstate distribution of the pump model in 103 waveguides with a long-long defect with various pump power at the first and last step of the evolution, and the pump light is injected at the site $-1$. 
  The initial pump power is set as (a)--(c)~\qty{1}{\watt}, (d)--(f)~\qty{30}{\watt}, and (g)--(i)~\qty{100}{\watt}. 
  (a),(d),(g)~The corresponding system spectrum. 
  (b),(e),(h)~The zero-energy, maximum-energy, and minimum-energy eigenmode distribution over the lattice at the first step of the evolution.
  (c),(f),(i)~The zero-energy, maximum-energy and minimum-energy eigenmode distribution over the lattice at the last step of the evolution.}
  \label{fig:eigenstate_longlong_pump}
\end{figure*}

The reason that the topological entangled biphoton weight increases is that the topology of the defect is converted from a long-long defect into novel defects by the large pump power. 
The high overlap between the topologically trivial localized eigenmode and the zero-energy topologically trivial eigenmode during the SFWM process is the key for the generation of biphotons. 
The pattern in the topological entangled biphoton weights shown in Fig.~\ref{fig:heatmap-weight-gap}(a) is 
affected by the following three factors:
first is the 
spectrum gap between the topological trivial maximum eigenmode and the bulk band,
as shown in
the heat map in
Fig.~\ref{fig:heatmap-weight-gap}(b);
the second is the overlap between the zero-energy topological eigenmode and the localized topological trivial maximum and minimum eigenmodes;
and the third is the overlap between the input state
and the trivial localized eigenmodes. 

The emergence of the high weight of topological biphoton states in the generated biphoton states
is the result of the satisfaction of the aforementioned three 
factors: in the spectrum, there exist energy-isolated trivial states that have nontrivial overlaps with both the input state and the topological zero-energy eigenmode.
As shown in Figs.~\ref{fig:eigenstate_longlong_pump}(e) and~\ref{fig:eigenstate_longlong_pump}(f), 
when
the pump power is \qty{30}{\watt}, 
the two trivial localized
modes have high amplitude at the input lattice (waveguide indexed $-1$), and the overlap between the 
topological zero-energy and the localized topological trivial eigenmodes have high overlap at waveguides indexed~$0$ and~$\pm 2$. 

As
we continue to
increase 
the pump power, the topology of the defect differs, and the shapes of the distributions of the topologically trivial localized eigenmode 
and the topological zero-energy eigenmode vary. 
The overlap between the 
topologically trivial localized eigenmodes and the topological zero-energy eigenmode decreases. 
The case with pump power \qty{100}{\watt} is shown in Figs.~\ref{fig:eigenstate_longlong_pump}(g)--\ref{fig:eigenstate_longlong_pump}(i). 
With the pump power of
\qty{100}{\watt}, 
the topologically protected edge state dominates at $\pm 2$ and $\pm4$ and exponentially decays into the bulk. 
Meanwhile, the trivial localized states mainly distribute at $0,\pm 1$ sites.
The overlap between the topological zero-energy eigenmode and the topological trivial localized eigenmodes is quite small. 
Compared with the case without
nonlinear mechanics
in 
Eq.~\eqref{eq:H-kerr},
the shift of the localization is the result of the expanded effective region of defects due to the nonlinear gain. 
Therefore, in the extremely large pump power region, the weight of topologically entangled biphoton states drops dramatically.  As seen from Fig.~\ref{fig:heatmap-weight-gap}(a), the defect topology transition happens around a pump power of \qty{35}{\watt} for our setup, which is the green dashed line therein.

The striped pattern in the region with high weight of topological biphoton state, as shown in Fig.~\ref{fig:heatmap-weight-gap}(a), results from the spectrum dynamics during the biphoton evolutions in Figs.~\ref{fig:eigenstate_longlong_pump}(d) and~\ref{fig:eigenstate_longlong_pump}(g), where, as the evolution proceeds, the spectrum gap between the maximum eigenmode and the bulk oscillates and decreases. Furthermore, the location consistency of the stripes in Figs.~\ref{fig:heatmap-weight-gap}(a) and~\ref{fig:heatmap-weight-gap}(b) also reflects this point. 

\subsection{Robustness to disorder}
Topological edge states exhibit inherent robustness against disorder and perturbations due to their topological protection, which arises from global symmetries rather than local material details. This resilience enables fault-tolerant applications in quantum computing, photonics, and energy-efficient electronics by preserving coherent transport even in imperfect environments.

\begin{figure*}
  \centering
(a)~\includegraphics[width=0.45\linewidth]{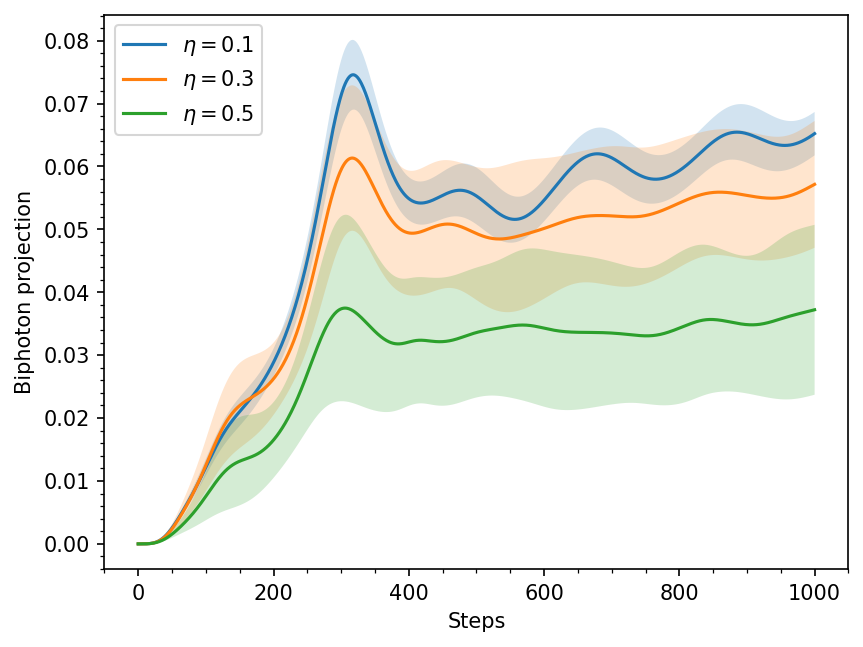}
(b)~\includegraphics[width=0.45\linewidth]{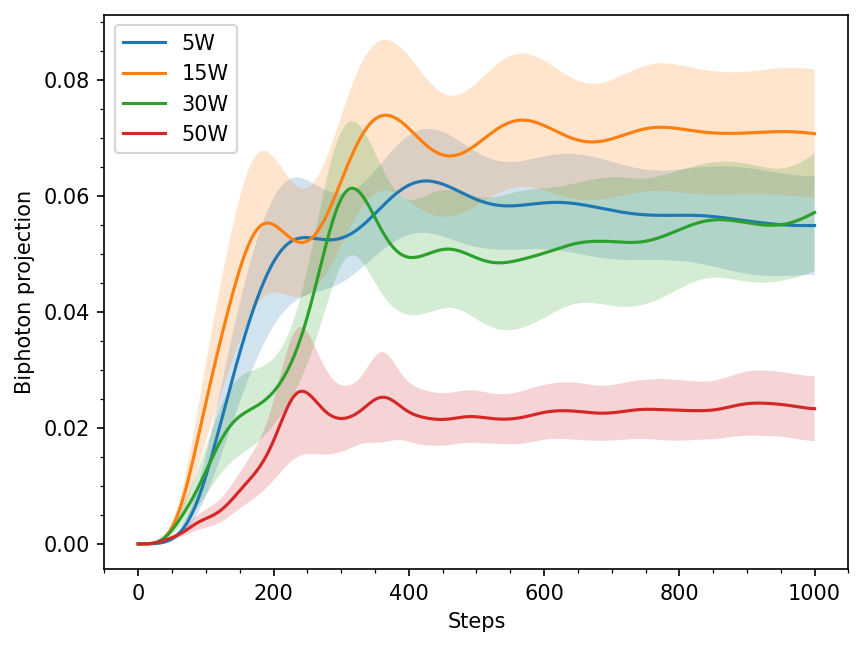}
  \caption{(a)~The dynamics of topological biphoton weight in the generated biphoton states during the system and evolution steps,
  with various disorder strengths $\eta=0.1$, 0.2, 0.3, 0.4, and 0.5, 
  and with a pump power of \qty{30}{\watt}, which is the pump value generating the high biphoton weight in the dynamics, as seen in Fig.~\ref{fig:heatmap-weight-gap}(a).
  (b)~Heat map of the topological biphoton weight in the generated biphoton states with pump powers of 5, 15, 30, and~\qty{50}{\watt}, 
and evolution steps, where the system disorder strength is set to $\eta=0.3$. For each disorder value $\eta$, we generate 20 corresponding disorder configurations to show the variations in panel~(a) and the average value in panel~(b).}
  \label{fig:weight-eta-step}
\end{figure*}

To investigate the robustness of our method, we artificially add waveguide fabrication disorder to investigate the robustness of our scheme. Specifically, the disordered waveguide couplings follow $[ v(1-\eta),v(1+\eta)]$, with $\eta\in[0,1]$ as the disorder strength and $v$ as the ideal coupling strength between waveguides, which is determined by the fabrication distance between waveguides. 
We investigate the effect of disorder on the weight of generated topologically entangled biphoton states.
As shown in Fig.~\ref{fig:weight-eta-step}(a), in the case with pump power \qty{30}{\watt}, with various disorder strengths $\eta$, we present the
average and the
variation in the weight of topologically entangled biphoton states during the evolutions. 
The weight of generated topological biphoton states drops as the system disorder strength increases, which is reasonable, and our simulations show that our method manages to ensure the generation of topological biphoton states even with a high disorder of $\eta=0.5$ with a weight around $3.5\%$. 
In all our simulations, the values for disorder cases are analyzed over 20 disordered configurations generated randomly.

To investigate the effect of disorder on the self-induced topology transitions in our setup, we study the topological biphoton weight in the generated biphoton states concerning the pump power and evolution steps. 
The study of the cases with disorder strength $\eta=0.3$ is shown in Fig.~\ref{fig:weight-eta-step} (b). We see that, with disorder strength $\eta=0.3$, the weight of topologically entangled biphotons in the generated biphoton states demonstrates similar behavior as in Fig.~\ref{fig:heatmap-weight-gap} (a). 
With the increase of pump power, the weight of the topological biphoton state increases first, then drops. With $\eta=0.3$, we find the highest weight of generated topological biphoton states in our setup is when the pump power is \qty{15}{\watt} [among the pump powers under investigation~(5, 15, 30, and~\qty{50}{\watt})], and the high weight does not decrease much and still holds at around 7\%, which demonstrates the robustness of our method.

\section{A photonic platform realization proposal}

In this section, we present a generalization of our scheme, where each coupling strength of the nearest-neighbor lattice is controllable via the on-site power. This generalization belongs to the 
contemporaneous ``active''
topological photonics platforms, such as the chip design in Ref.~\cite{PhysRevApplied.22.054011}, where the coupling in all lattice sites is controllable and requires an exquisite global control strategy. However, an actively controlled nonlinearly coupled waveguide with SFWM has not yet been reported experimentally to our knowledge. The photonic time-bin platform has demonstrated its versatility in many physical simulation scenarios, such as the observation of topological edge states~\cite{PhysRevLett.121.100502}, characterization of a photonic quantum anomalous Hall system~\cite{PhysRevLett.131.133601}, and observation of the non-Hermitian edge burst effect~\cite{PhysRevLett.132.203801}. Thus, the photonic time-bin platform is a good candidate for realizing our generalized scheme with the involvement of `active' control components. 

\begin{figure}
    \centering
    \includegraphics[width=0.5\textwidth]{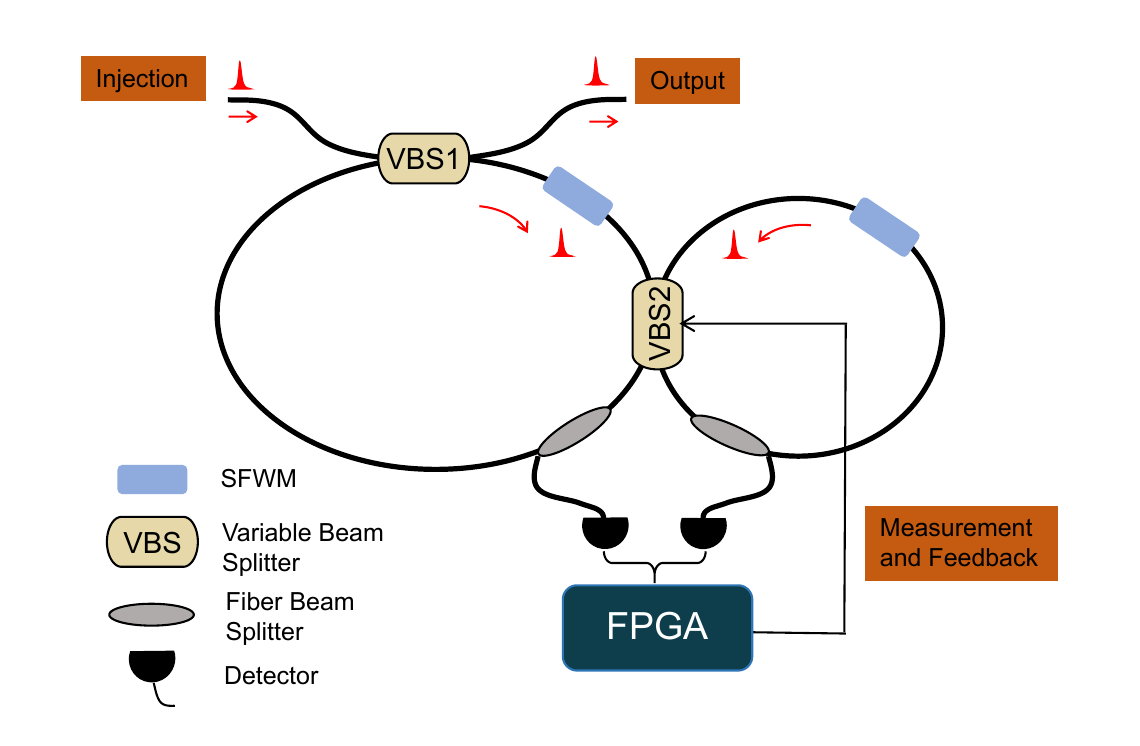}
    \caption{Actively controlled scheme with time-bin encoding. The lattice sites in the SSH model are encoded by the time bins traveling in the fiber loops. Spontaneous four-wave mixing occurs when the pump pulses go through the embedded nonlinear materials. A slight proportion of the pump pulses is split out for measurement, and the results are fed into a field-programmable gate array to compute the signals to change the transmission reflection ratio of VBS2. VBS1 is used to couple the pump laser in and the biphoton out of the fiber loops.}
    \label{fig:time-bin-exp}
\end{figure}

Our time-bin realization setup is shown in Fig.~\ref{fig:time-bin-exp}, which is mainly composed of two fiber loops. One represents sublattice A and the other represents sublattice B in the SSH model. The initial laser pulse is injected into the fiber loops by a variable beam splitter~(VBS1) and goes through the nonlinear material where entangled biphotons are generated. Both the pump laser pulse and the biphotons evolve step by step in the fiber loops. Every two adjacent pulses has a fixed time delay due to the length difference of the two fiber loops, and their coupling can be controlled independently via VBS2. To simulate the pump-dependent nonlinear coupling effect, a slight proportion of the pump power is split out for measurement in each round trip. Using the measurement results, a field-programmable gate array~(FPGA) computes the feedback signal for VBS2 to tune the transmission reflection ratio, which corresponds to adjusting the coupling of each site for the next step. The generated biphotons can be output in any step by switching VBS1 to reflecting mode.

With current technologies in time-bin experiments~\cite{mcmahonFullyProgrammable100spin2016,hamerlyExperimentalInvestigationPerformance2019},  our experimental proposal is accessible to demonstrate our results. Specifically,  for a pump power around $\qty{30}{W}$ in our theoretical analysis, which is found to be a suitable power setting for the demonstration of the aforementioned nonlinearity effect in our work,  the pump laser is suggested to be set as a picosecond pulsed laser~(\qty{10}{ps}) with a repetition frequency of \qty{100}{MHz} and laser power of~\qty{30}{mW}.   
The length of the fiber loops is suggested to be set as about~\qty{300}{m}, so that each round trip is able to contain more than~100 pulses and the circuit has enough time to receive and respond to the feedback signal from the FPGA.

Our time-bin encoded platform is more flexible and programmable, which can simulate not only pump-controlled nonlinear coupling but also its generalizations of the `active' modular system mentioned above. With our setup, the shift of the position of the defect is manageable, enabling the topological biphoton to be coupled out in an arbitrarily appointed time bin. 
In another way, the waveguide scheme features advances in its direct compatibility with current Mach-Zehnder-type or microring-type photonic chips~\cite{daiProgrammableTopologicalPhotonic2024,PhysRevApplied.22.054011}. An on-chip plug-in device based on the silicon waveguide scheme can be designed to provide an entangled biphoton source for the following integrated large-scale silicon-based photonic chips. Although challenging, recent research facilitates the integration of time-bin systems with photonic integrated chips for possible applications across quantum information and topological technologies~\cite{xiongCompactReconfigurableSilicon2015,fangOnchipGenerationTimeand2018}. 

\section{Outlook and Conclusion}
In this work, we present a flexible silicon waveguide chip design for the manipulation of topological biphoton states. By including the nonlinear gain and SFWM mechanics, the topology of a chip defect can be manipulated with the external pump power. 
Topology-driven localization further suppresses disorder-induced scattering of the generated topologically entangled biphotons. Therefore, our scheme is robust to system disorder and enables the reusability of silicon waveguide chips.
We also present a time-bin realization for our scheme and its generalization of a contemporaneous ``active'' module, which demonstrates the versatility of our method.

This externally reconfigurable chip enables fault-tolerant quantum state manipulation for integrated photonic quantum networks. Its topological protection allows noise-resilient quantum logic gates and quantum memory in multicore waveguide arrays. The pump-controlled defect modes support programmable quantum routing for photonic cluster states in quantum computing architectures. Additionally, the CMOS-compatible design facilitates scalable production of quantum light sources and topological quantum repeaters, essential for metropolitan-scale quantum communication. System reusability and intrinsic robustness against manufacturing variations reduce operational costs, accelerating the industrial deployment of topology-enhanced quantum photonic circuits. 
This design bridges nonlinear topology and quantum photonics, offering a manufacturable solution for robust quantum information processing, which has promoted the industrialization of quantum technology.

\vspace{1cm}

\section*{Acknowledgements} \label{sec:acknowledgements}
  We acknowledge support from the National Natural Science Foundation of China~(Grants No.~62571434, No.~12104101, and No.~12304398), from the Fundamental Research Funds for the Central Universities,
   China Postdoctoral Science Foundation~(Grant No.~2023M731532), and from Ningbo Yongjiang Talent Program~(Grant No.~2024A-366-G).
  J.W. would like to thank Yucheng Wang and Wei Sun for valuable discussions.

\bibliography{reference}

\end{document}